\documentclass[twocolumn,showpacs,preprintnumbers,
                amsmath,amssymb,eqsecnum,aps]{revtex4}
\usepackage{epsfig}
\usepackage{graphicx}
\usepackage{dcolumn}
\usepackage{bm}

\begin{document}



      \title{
          Semiclassical approach to the decay of protons in circular motion
             under the influence of gravitational fields }


\author{Douglas Fregolente}
\email{douglas@ift.unesp.br}
\author{George E.\ A.\ Matsas}
\email{matsas@ift.unesp.br}
\affiliation{Instituto de F\'\i sica Te\'orica,
             Universidade Estadual Paulista,
             R. Pamplona 145, 01405-900, S\~ao Paulo,
             S\~ao Paulo, Brazil
             }
\author{Daniel A.\ T.\ Vanzella}
\email{vanzella@ifsc.usp.br}
\affiliation{Instituto de F\'\i sica de S\~ao Carlos,
             Universidade de S\~ao Paulo,
             Cx. Postal 369,
             13560-970, S\~ao Carlos, S\~ao Paulo, Brazil}



\begin{abstract}

We investigate the possible decay of protons in geodesic circular motion
around neutral compact objects. Weak and strong decay rates and the
associated emitted powers are calculated using a semi-classical approach.
Our results are discussed with respect to distinct ones in the literature,
which consider the decay of accelerated protons in electromagnetic fields.
A number of consistency checks are presented along the paper.

\end{abstract}


\maketitle


\section{\label{intro} Introduction}

It is well known that according to the particle standard model inertial
protons are stable. However, this is not so if the proton is under the
influence of some external force because in this case the accelerating
agent can provide the required extra energy, which allows the proton to
decay. To the best of our present knowledge, the first ones to consider
the decay of accelerated protons and similar processes as
\begin{equation}
p^+ \stackrel{a}{\to} p^+\, \pi^0 \,,
\label{pppi0}
\end{equation}
were Ginzburg and Zharkov~\cite{GZ}-\cite{Zhar}. In Ref.~\cite{GZ}
the proton is described by a classical current with a well defined
trajectory while the pion is field quantized. This approach is
accurate in the no-recoil regime, i.e. when the relevant parameter
$\chi\equiv a/m_p$ involving the $p^+$'s proper acceleration $a$ and
mass $m_p$ is less than unity. At the same time, Zharkov~\cite{Zhar}
(see also Ref.~\cite{R} for a recent review) investigated the
process
\begin{equation}
p^+ \stackrel{A_\mu}{\to} p^+\, \pi^0
\label{pppi0Amu}
\end{equation}
and the strong and weak proton decays
\begin{eqnarray}
p^+ &\stackrel{A_\mu}{\to}&  n^0\, \pi^+ \,,
\label{pnpi+Amu}
\\
p^+ &\stackrel{A_\mu}{\to}& n^0\, e^+ \, \nu \,,
\label{pne+nuAmu}
\end{eqnarray}
respectively, in the presence of an electromagnetic field $A_\mu$,
where all particles are field quantized. For this purpose, it was
used the comprehensive formalism developed by  Nikishov and
Ritus~\cite{NR} (see also \cite{ritus69}), which allows one to
investigate quantum processes in such a background.  The study of
particle processes in the presence of strong electromagnetic fields
should be important to the analysis of certain aspects of high
energy cosmic ray physics produced in pulsars and magnetars. In such
intense magnetic fields ($H \sim 10^{12} - 10^{17}$ G) the strong
coupling of  protons and neutrons with mesons can generate $\rho$'s
and $\pi$'s with a non negligible intensity~\cite{TK}-\cite{berez}.

\begin{figure}[b]
\includegraphics[width=0.2\textwidth]{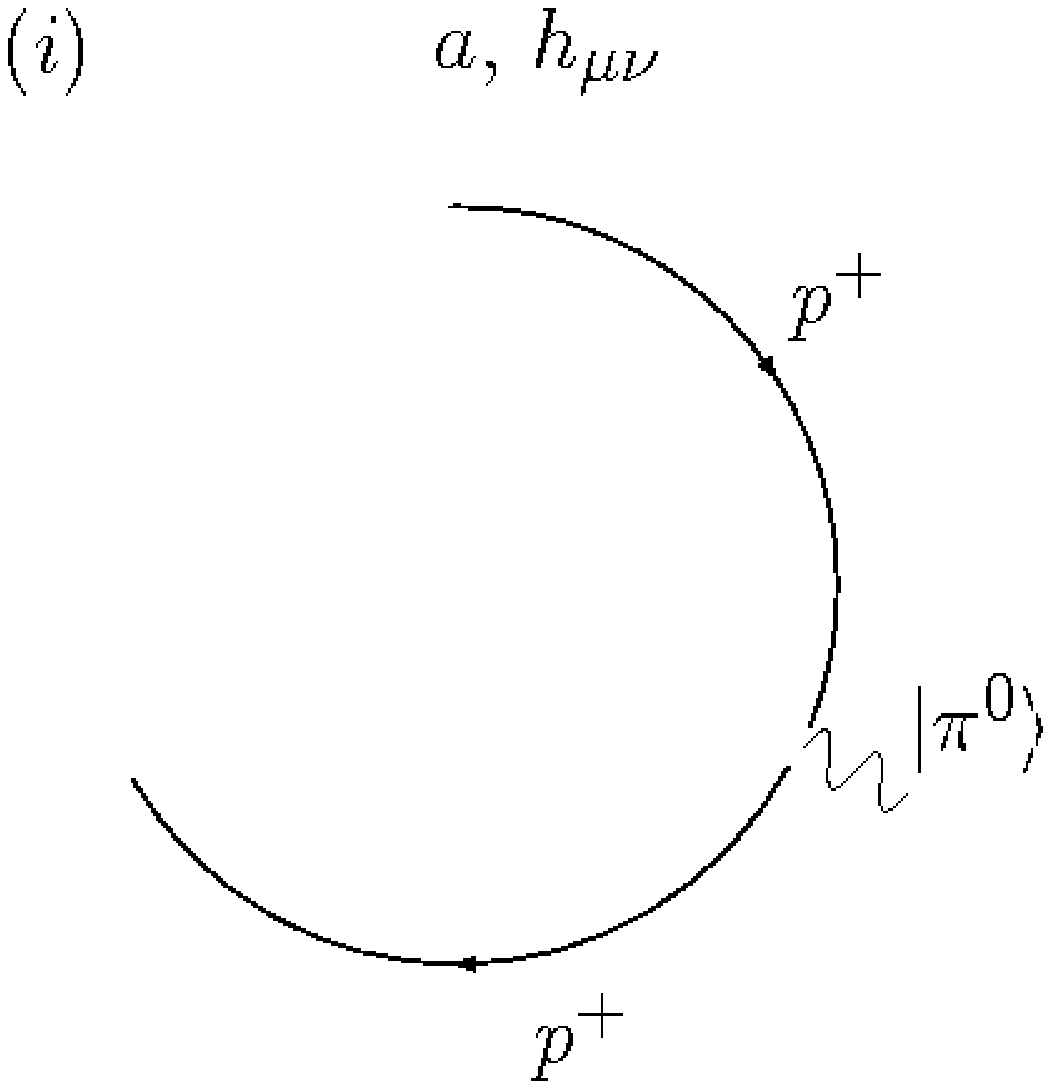}
\includegraphics[width=0.2\textwidth]{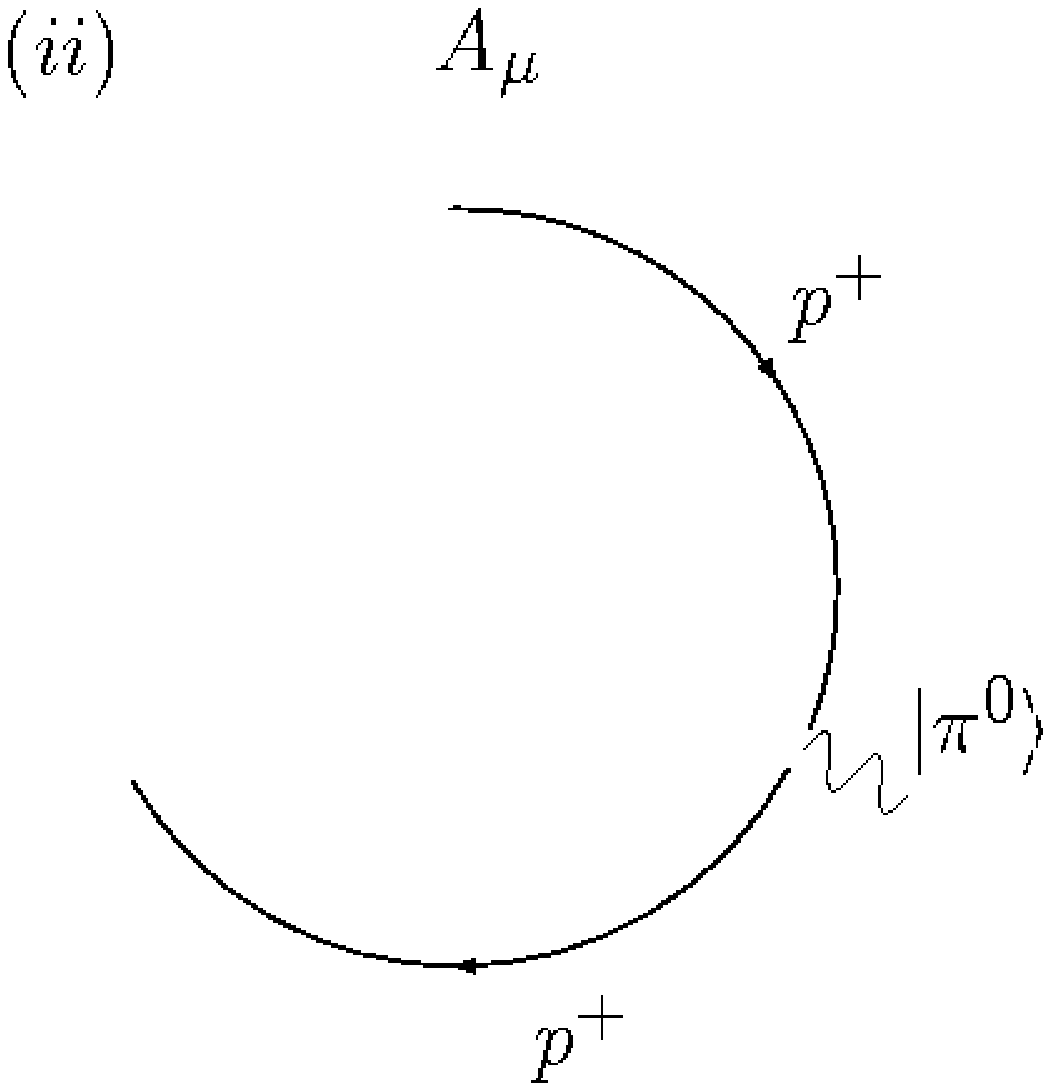}
\caption{\label{figpppi0} The process $p^+ \to p^+\, \pi^0$ is represented
in the presence of background gravitational ($h_{\mu \nu}$) and
electromagnetic ($A^\mu$) fields. In the proper regime, this
process can be well described in both backgrounds by the classical
current approach, where the proton is assumed to have a well defined
worldline with acceleration $a$ and the pion is field quantized.
Notice the similarity of the proton behavior in $(i)$ and $(ii)$.}
\end{figure}

\begin{figure}[t]
\includegraphics[width=0.2\textwidth]{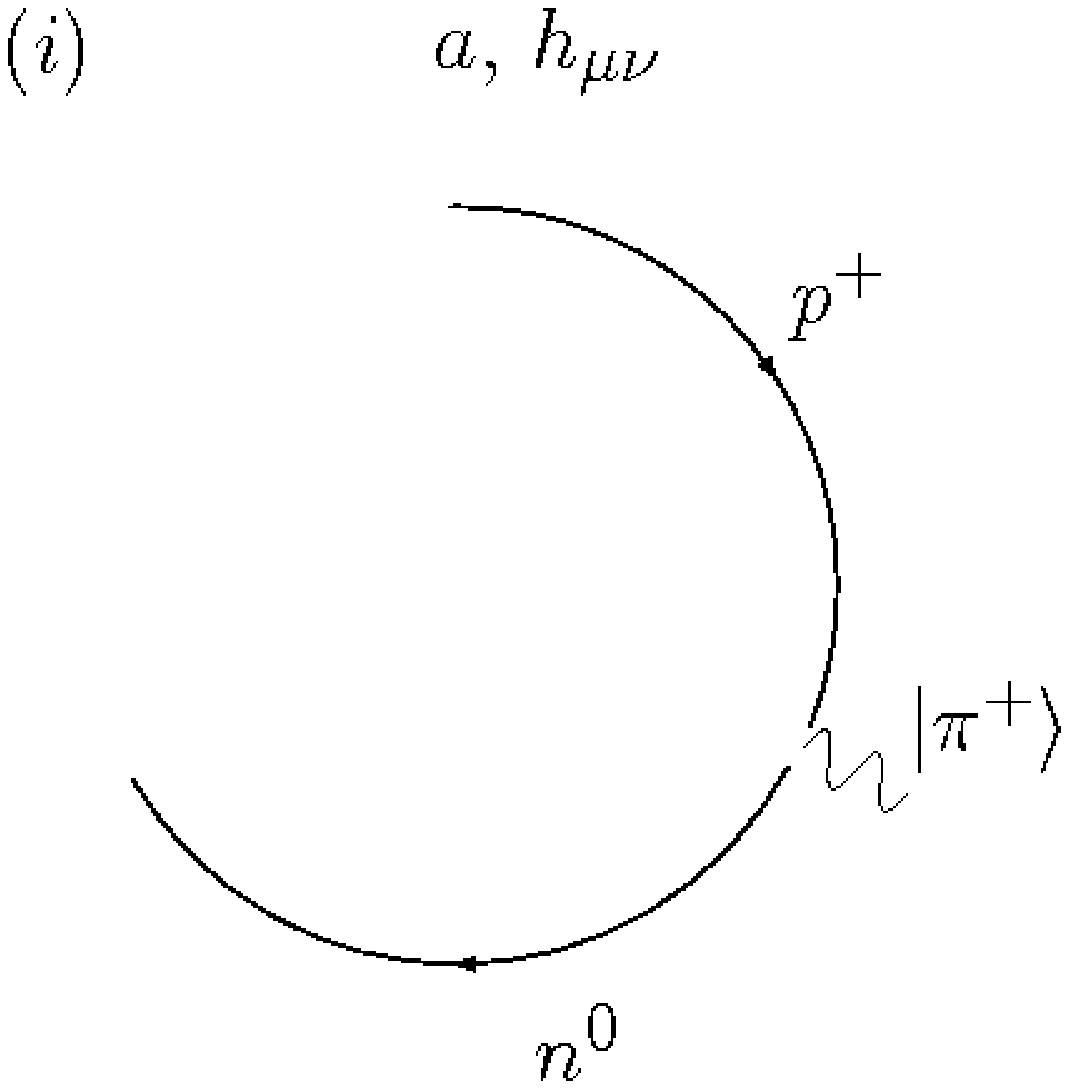}
\includegraphics[width=0.2\textwidth]{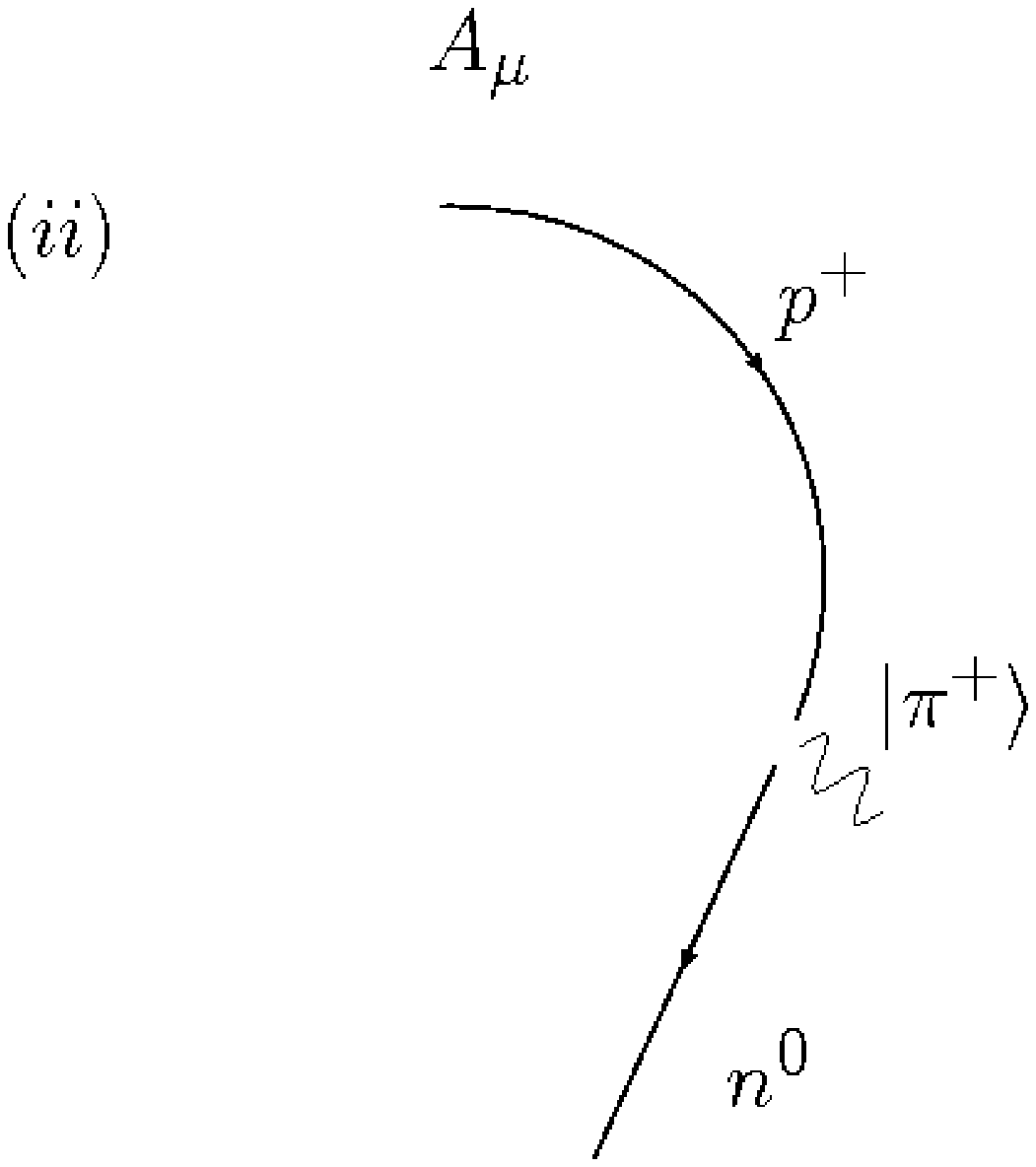}
\caption{\label{figpnpi+} The process $p^+ \to n^0\, \pi^+$
is represented in the presence of background gravitational ($h_{\mu \nu}$)
and electromagnetic ($A^\mu$) fields. Only in the first case it can be well
described, in the proper regime, by the classical current approach labeled
by the nucleons' proper acceleration $a$. Notice the difference of the
proton-neutron behavior in $(i)$ and $(ii)$.}
\end{figure}

\begin{figure}[t]
\includegraphics[width=0.2\textwidth]{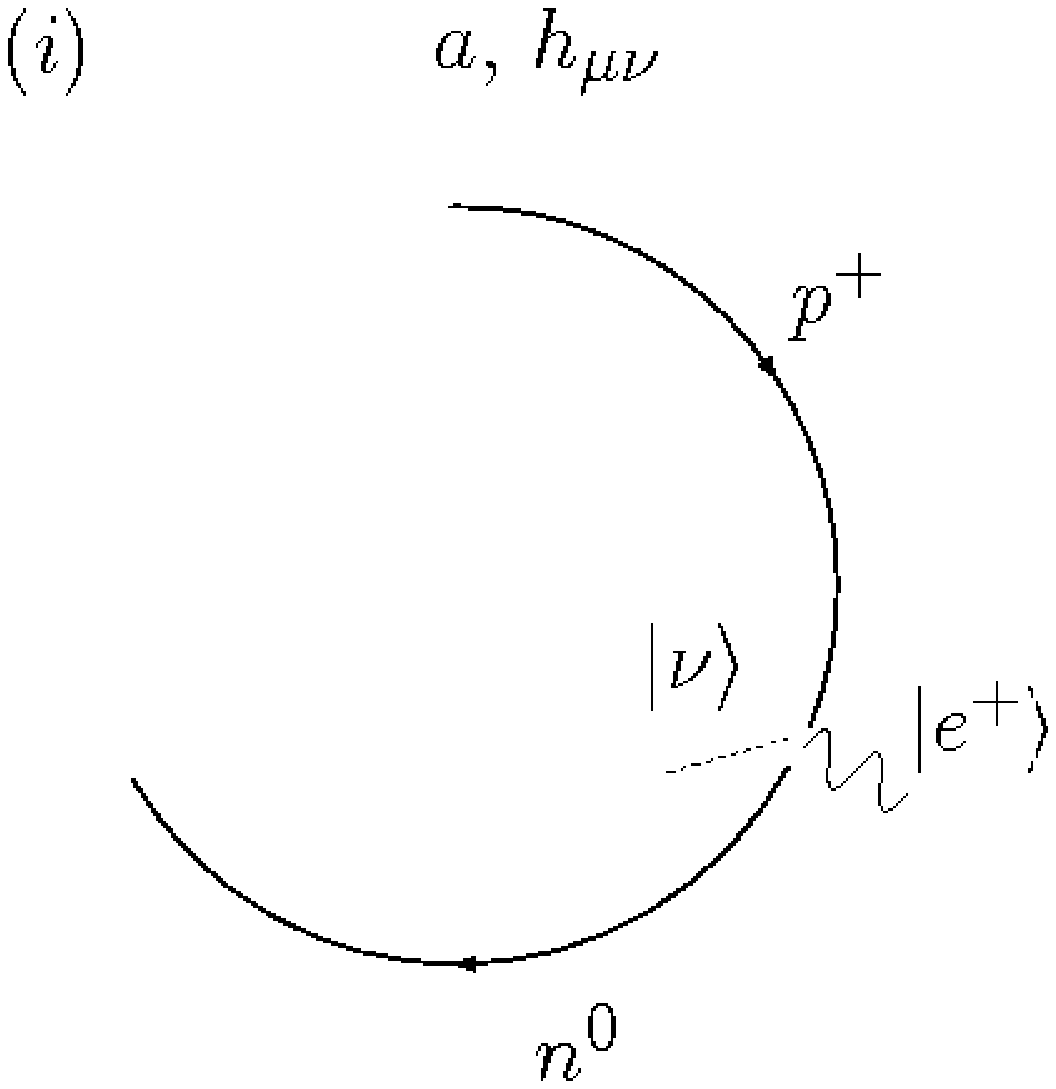}
\includegraphics[width=0.2\textwidth]{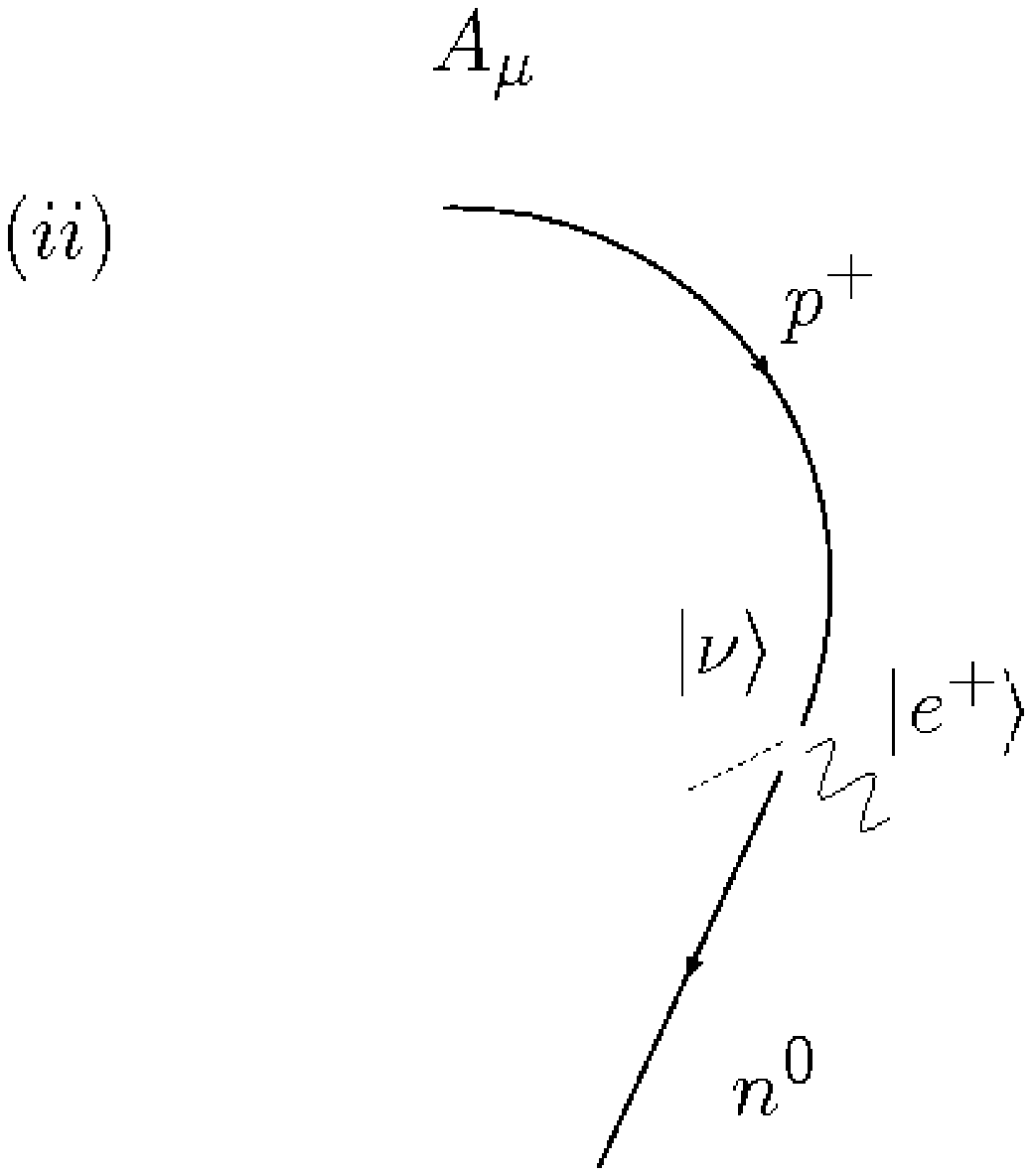}
\caption{\label{figweak} The process $p^+ \to n^0\, e^+ \, \nu$
is represented in the presence of background
gravitational ($h_{\mu \nu}$) and electromagnetic ($A^\mu$)
fields. Analogously to the previous figure, only the first case
can be well described, in the proper regime, by the classical current
approach.}
\end{figure}

In the proper regime (i.e. where backreaction effects are not important),
the reaction rate associated with processes~(\ref{pppi0}) when the $p^+$
is in circular motion and~(\ref{pppi0Amu}) when it is under the influence
of a magnetic field coincide. Notwithstanding, this is not so for the
processes~(\ref{pnpi+Amu})-(\ref{pne+nuAmu}), and
\begin{eqnarray}
p^+ &\stackrel{a}{\to}& n^0\, \pi^+ \,,
\label{pnpi+}
\\
p^+ &\stackrel{a}{\to}& n^0\, e^+ \, \nu \,,
\label{pne+nu2}
\end{eqnarray}
respectively (where the $p^+-n^0$ are described by a classical current).
This is a consequence of the fact that in the classical current approach
both, proton and neutron, are usually assumed to follow the same trajectory
in contrast with what really happens in the presence of a background
magnetic field (notice the difference between Fig.~\ref{figpppi0},
and Figs.~\ref{figpnpi+} and~\ref{figweak}).

This raises the question about what is the physical situation
simulated by the classical current method when applied to the
processes (\ref{pnpi+}) and (\ref{pne+nu2}), where one considers
that only mesons and leptons are field quantized. Once protons and
neutrons are described by a common current, one should look for a
situation where they are mainly undistinguishable. This is what
happens in gravitational fields according to the equivalence
principle. (For early and recent investigations on geodesic
synchrotron radiation from classical currents see
Refs.~\cite{Misneretal} and~\cite{Lemosetal}, respectively.) As a
consequence, processes~(\ref{pnpi+}) and~(\ref{pne+nu2}) should
represent fairly well the strong and weak conversion of protons into
neutrons when they orbit chargeless compact objects provided that
the back reaction on the neutron is negligible. This is what we are
going to investigate in this paper.

In our procedure, we take into account the proton-neutron mass difference,
by introducing a {\em semiclassical} rather than {\em classical} current.
We will be following Ref.~\cite{MV}, where a semiclassical current was
successfully used to model the decay of {\em linearly} accelerated protons
in the study of the Fulling-Davies-Unruh effect~\cite{FDU}.  The current is
``classical" in the sense that the proton-neutron is associated with a well
defined worldline and ``quantum" in the sense that it behaves as a two-level
quantum system.
(A simplified related calculation, where all particles are treated as
scalars can be found in Ref.~\cite{M}.) The calculation is performed
in Minkowski spacetime and the gravitational field is described by a
Newtonian-like central force.

There is also an important difference concerning the proton decay when it
is under the influence of a gravitational field rather than of a magnetic one,
which is worthwhile to call the attention. The physical scale for the proton
decay is given by its proper acceleration $a$. Because of the pion mass,
process~(\ref{pne+nu2}) dominates over process~(\ref{pnpi+}) in the region
$m_e + \Delta \mu < a < m_\pi + \Delta \mu$
and because of the magnitude of the strong coupling constant,
process~(\ref{pnpi+}) dominates over process~(\ref{pne+nu2}) in the
region $a > m_\pi + \Delta \mu$, where $\Delta \mu = m_n - m_p$. Now,
in the presence of a magnetic field $H$, the proper acceleration of a
proton in circular motion can be written as $a = \gamma e H /m_p$,
where $\gamma = E/m_p$ is the usual relativistic factor given by the
ratio of the proton's energy and mass. In the region $a > m_\pi + \Delta \mu$,
the strong process dominates over electromagnetic processes and energy
degradation through photon emission does not play any relevant role. This
is not so, however, in the region
$m_e + \Delta \mu < a < m_\pi + \Delta \mu$,
where electromagnetic processes dominate over the weak one and much of
the proton's energy $E$ can be carried away by the photons, driving its
acceleration  below the threshold $m_e + \Delta \mu$. (Recall that $\gamma$
is proportional to $E$.) The situation is quite different in a gravitational
field. Assuming Minkowski space, the proper acceleration
$a = R \Omega^2 \gamma^2 $ of a proton in circular orbit with radius $R$
and angular velocity $\Omega$ around a compact object with mass $M$ can
be written as
$a= (GM \Omega^4)^{1/3}/( 1-(GM\Omega)^{2/3})$,
where we have used the Newtonian gravity relation $R^3 \Omega^2 = GM$.
Then, as the orbiting proton emits photons descending to a more internal
orbit with larger $\Omega$, its proper acceleration tends to increase
rather than to decrease, in contrast to the electromagnetic case.
Whether or not a proton decays along its inspiraling trajectory will
depend on the mass of the central object and other details, which will
be discussed further.

Now one may wonder how accurate can be our results when applied to quite
strong gravitational fields. In principle, an exact calculation would
require that we take into account the spacetime curvature in the particle
field quantization. However, as it was shown in Refs.~\cite{CHM}
and~\cite{CCMM} the results obtained assuming a full curved spacetime
ruled by Einstein equations and a flat background endowed with a Newtonian
attraction force  should not differ by more than 20\%-30\% up to the
inner {\em stable} circular orbit of a static black hole. This is going
to suffice for our purposes.

The paper is  organized as follows: In Sec.~\ref{formalism} we present
the semiclassical current formalism. In Sec.~\ref{current} we evaluate
the scalar and fermion emission rates for a uniformly swirling current.
In Sec.~\ref{power} we evaluate the corresponding radiated powers.
In Sec.~\ref{proton} we use the previous results to analyze the decay
of protons orbiting chargeless compact objects. Consistency checks for
our formulas are presented. We dedicate Sec.~\ref{discussion} to our final
discussions. We assume Minkowski spacetime with metric components
$\eta_{\mu \nu} = {\rm diag} (+1, -1, -1, -1)$
associated with the usual inertial coordinates
$(t, {\bf x})$ and adopt natural units $c = \hbar = 1$ throughout this
paper unless stated otherwise.


\section{\label{formalism} Semiclassical current formalism}

Let us consider the following class of processes
\begin{eqnarray}
p_1 \; &\to& \; p_2 \;  g \;,
\label{p1p2a}\\
p_1 \; &\to& \; p_2 \;  f_1 \;\bar f_2 \;,
\label{p1p2b}
\end{eqnarray}
where a scalar $g$ or a fermion-antifermion pair $f_1$-$\bar f_2$ are
emitted as the particle $p_1$ evolves into $p_2$.
The $g$, $f_1$, $\bar f_2$, $p_1$ and $p_2$'s rest masses are
$m$, $m_1$, $m_2$, $M_1$ and $M_2$,  respectively. We will be
interested here in cases where $m, m_1, m_2 \ll M_1, M_2$. The
particle emission will be assumed not to change significantly
the four-velocity of $p_2$ with respect to $p_1$. This is called
{\em ``no-recoil condition''}, which is verified when the momentum
of the emitted particles with respect to the instantaneously inertial
{\em rest frame} lying at $p_1$ satisfies $|{\bf k}_{\rm rf}| \ll M_1,\;M_2$.
Because $m, m_1, m_2 \ll M_1, M_2$, this implies that the energy of each
emitted particle satisfies ${\omega}_{\rm rf}  \ll M_1,\;M_2$. As the
typical energy $\omega_{\rm rf}$ of the emitted particles is of the order
of $p_1$'s {\em proper} acceleration $ a \equiv \sqrt{|a_\mu a^\mu|}$,
the no-recoil condition can be recast in the {\em frame independent}
form~\cite{VM3}
\begin{equation}
a \ll M_1,\; M_2 \;.
\label{norecoilcondition}
\end{equation}

The particles $p_1$ and $p_2$ will be seen as distinct energy eigenstates
$\vert p_1 \rangle$ and $\vert p_2 \rangle$, respectively, of a two-level
system. The associated {\it proper} Hamiltonian $\hat H_0$ of the particle
system satisfies, thus,
\begin{equation}
\hat H_0 \; \vert p_j \rangle = M_j \; \vert p_j \rangle\;,\;\; j=1,2\;.
\label{H0}
\end{equation}
We shall describe our pointlike particle system $p_1$-$p_2$
in the process~(\ref{p1p2a}) by the semiclassical {\em scalar}
source
\begin{equation}
\hat j(x)=
[\hat q(\tau)/u^0 (\tau)]\; \delta^3 [{\bf x}-{\bf x}(\tau)]
\label{CIa}
\end{equation}
and in the process~(\ref{p1p2b}) by the {\em vector} current
\begin{equation}
\hat j^\mu (x)=
[\hat q(\tau)\;u^\mu (\tau)/u^0 (\tau)] \; \delta^3 [{\bf x}-{\bf x}(\tau)]\;.
\label{CI}
\end{equation}
Here $x^\mu(\tau)$ is the classical world line parametrized by the proper
time $\tau$ associated with $p_1$-$p_2$, $u^\mu (\tau) \equiv dx^\mu/d\tau $
is the corresponding four-velocity, and
$\hat q (\tau ) \equiv e^{i \hat H_0 \tau} \hat q_0 e^{-i \hat H_0 \tau}$,
where $\hat q_0$ is a self-adjoint operator evolved by the one-parameter
group of unitary operators $ \hat U (\tau ) = e^{-i\hat H_0 \tau}$.

The emitted scalar $g$ in the process~(\ref{p1p2a}) is associated with a
complex Klein-Gordon field
\begin{equation}
\hat \Phi(x)=
             \int d^3 {\bf k}
\left[
     \hat a_{{\bf k} } \phi^{(+\omega)}_{{\bf k} } (x)
     + \hat c^\dagger_{{\bf k} } \phi^{(-\omega)}_{-{\bf k} } (x)
\right]\;,
\label{KGF}
\end{equation}
while the emitted fermions $f_1- \bar f_2$ in the process~(\ref{p1p2b})
are associated with the fermionic one
\begin{equation}
\hat \Psi_i (x)=
             \sum_{\sigma = \pm } \int d^3 {\bf k}_i
\left[
     \hat b_{{\bf k}_i \sigma} \psi^{(+\omega_i)}_{{\bf k}_i \sigma} (x)
     + \hat d^\dagger_{{\bf k}_i \sigma} \psi^{(-\omega_i)}_{-{\bf k}_i
     -\sigma} (x)
\right]\;,
\label{FF}
\end{equation}
where $i=1,2$ labels the two fermions.
Here
$ \hat a_{{\bf k}} \;( \hat a^\dagger_{{\bf k}})$,
$ \hat b_{{\bf k}_i \sigma} \;(\hat b^\dagger_{{\bf k}_i \sigma})$,
$ \hat c_{{\bf k}} \;( \hat c^\dagger_{{\bf k}})$
and
$ \hat d_{{\bf k}_i \sigma} \;(\hat d^\dagger_{{\bf k}_i \sigma})$
are annihilation (creation) operators of scalars, fermions, antiscalars
and antifermions, respectively, with three-momentum ${\bf k}=(k^x,k^y,k^z)$
and energy
$ \omega=\sqrt{{\bf k}^2+m^2}$
for the scalar, and
${\bf k}_i=(k_i^x,k_i^y,k_i^z)$
and
$ \omega_i=\sqrt{{\bf k}_i^2+m_i^2}$
for the fermions.
$ \phi^{(\pm \omega)}_{\bf k}$
and
$ \psi^{(\pm \omega_i)}_{{\bf k}_i \sigma}$
are positive $[(+\omega), (+\omega_i)]$ and negative
$[(-\omega), (-\omega_i)]$ frequency solutions of the Klein-Gordon
$(\Box - m^2) \phi^{(\pm \omega)}_{\bf k} =0$
and Dirac
$(i\gamma^\mu \partial_\mu-m_i) \psi^{(\pm \omega_i)}_{{\bf k}_i\sigma}=0$
equations, respectively, where $\sigma$ labels the fermion polarization.

Next, we minimally couple the fields to our semiclassical source~(\ref{CIa})
and current~(\ref{CI}) according to the actions~\cite{IZ}-\cite{CQ}
\begin{equation}
\hat S_I^{(s)}  =
                  \int d^4x \;\hat j(x)
                  \left[
                         \hat{\Phi}(x)+ \hat{\Phi}^\dagger(x)
                  \right]
\label{Ss}
\end{equation}
for the scalar and
\begin{eqnarray}
\hat S_I^{(f)} &=& 
           \int d^4x\; \hat j_\mu(\hat{\bar \Psi}_1 \gamma^\mu (c_V-c_A\gamma^5) \hat \Psi_2 
\nonumber \\
	         &+& 
	         \hat{\bar \Psi}_2 \gamma^\mu (c_V-c_A\gamma^5)\hat \Psi_1 ) 
\label{Sf}
\end{eqnarray}
for the fermionic cases, (\ref{p1p2a}) and (\ref{p1p2b}),
respectively, where $\hat{\bar \Psi} = \hat{\Psi^\dagger} \gamma^0$
and $c_V=c_A=1$ in the processes here analyzed.

The  transition amplitudes at the tree level for the processes~(\ref{p1p2a})
and~(\ref{p1p2b}) are given by
\begin{equation}
{\cal A}_{{\bf k}} =
\; \langle p_2  \vert \otimes \langle {g}_{{\bf k}} \vert \;
\hat S_I^{(s)} \;
\vert  0 \rangle \otimes \vert p_1 \rangle \; ,
\label{AMPs}
\end{equation}
and
\begin{equation}
{\cal A}^{\sigma_1 \sigma_2}_{{\bf k}_1 {\bf k}_2} =
\; \langle  p_2
\vert \otimes \langle {f_1}_{{\bf k}_1 \sigma_1} ,
\bar {f_2}_{{\bf k}_2 \sigma_2}
\vert \;
\hat S_I^{(f)} \;
\vert 0 \rangle \otimes
\vert p_1  \rangle \; ,
\label{AMPf}
\end{equation}
respectively.
The differential transition probabilities are
\begin{eqnarray}
\frac{d{\cal P}_s^{p_1 \to p_2}}{d^3{\bf k}}
&=& \vert {\cal A}_{{\bf k}} \vert^2
\nonumber \\
&=& \frac{ G_{\rm eff}^{(s) 2}}{2 (2\pi)^3 \omega}
\int_{-\infty}^{+\infty} \!\!\!\! d\tau
\int_{-\infty}^{+\infty} \!\!\!\! d\tau'
\exp \{ i \Delta \mu (\tau-\tau')
\nonumber \\
& &
+ ik^\lambda [x(\tau)-x(\tau')]_\lambda
        \}
\label{dP3a}
\end{eqnarray}
and
\begin{eqnarray}
\frac{d{\cal P}_f^{p_1 \to p_2}}{d^3{\bf k}_1 d^3{\bf k}_2}
&=&
\sum_{\sigma_1=\pm} \sum_{\sigma_2=\pm} \vert
{\cal A}^{\sigma_1 \sigma_2}_{{\bf k}_1 {\bf k}_2} \vert^2
\nonumber \\
&=& \!\!\! \frac{2 \; G_{\rm eff}^{(f) 2}}{(2\pi)^6 \omega_1 \omega_2}\!
\int_{-\infty}^{+\infty} \!\!\!\!\! d\tau
\int_{-\infty}^{+\infty} \!\!\!\!\! d\tau'
\exp\{i\Delta \mu (\tau-\tau')
\nonumber \\
& +& i(k_1+k_2)^\lambda [x(\tau)-x(\tau')]_\lambda\}
\nonumber \\
&\times & \!\!\! \{ 2 k_1^{(\mu} k_2^{\nu)}  u_\mu(\tau)u_\nu(\tau')
            - k_1^\alpha {k_2}_\alpha  u^\mu(\tau) u_\mu(\tau')
\nonumber \\
& +& i \epsilon^{\mu\nu\alpha\beta} k_{1\alpha} k_{2\beta}
         u_\mu(\tau)u_\nu(\tau')
       \} \;,
\label{dP3}
\end{eqnarray}
accordingly,
where $\epsilon^{\mu\alpha\nu\beta}$ is the totally skew-symmetric
Levi-Civita pseudo-tensor (with $\epsilon^{0123}=-1$),
$ k_1^{(\mu}k_2^{\nu)} \equiv (k_1^{\mu}k_2^{\nu}+k_1^{\nu}k_2^{\mu})/2$,
$\Delta \mu \equiv M_2 - M_1$
and
$ G_{\rm eff}^{(s),(f)} \equiv
\vert \langle p_2 \vert \hat q_0 \vert p_1 \rangle \vert$
are the effective coupling constants for the scalar $(s)$
and fermionic $(f)$ channels.


\section{\label{current} Emission rates}

The world line of a  particle with uniform circular motion
with radius $R$ and angular velocity $\Omega$ as defined
by laboratory observers at rest in an inertial frame with
coordinates $(t,{\bf x})$, is
\begin{equation}
x^\mu(\tau)= (t\;,\;R \cos (\Omega t)\;,\;R \sin (\Omega t)\;,\; 0)
\label{WL}
\end{equation}
and the corresponding four-velocity is
\begin{equation}
u^\mu(\tau)= \gamma \,
            (1\;,\;
             - R \Omega \sin (\Omega t)\;,\;
             R \Omega \cos (\Omega t) \;,\;
             0)\;,
\label{UACC}
\end{equation}
where
$\gamma \equiv (1-R^2 \Omega^2)^{-1/2} =  \rm const$
is the Lorentz factor ($v \equiv R \Omega < 1$),
$t=\gamma \tau$,
and
$ a = \sqrt{-a_\mu a^\mu} = R \, \Omega^2 \gamma^2$
is the proper acceleration. Let us calculate now separately
the scalar and fermionic emission rates associated with
processes~(\ref{p1p2a}) and (\ref{p1p2b}), respectively.

\subsection{\label{scalar case} Scalar case}

First, let us analyze the process (\ref{p1p2a}). In order to decouple
the integrals in Eq.~(\ref{dP3a}), we define new coordinates,
\begin{equation}
\sigma \equiv \gamma ({\tau-\tau '})/{2}\;\;\;\;{\rm and}\;\;\;\;
s \equiv \gamma({\tau+\tau '})/{2}\;,
\label{CCGs}
\end{equation}
and perform the change in the momentum variable
\begin{equation}
k^\mu \mapsto {\widetilde{k}}^\mu
=
(
 {\widetilde{\omega}}\; , {\widetilde{\bf k}}
)\;,
\label{Gamma/k1k2s}
\end{equation}
where
\begin{eqnarray}
{\widetilde{\omega}} \; &=& \omega \; ,
\nonumber \\
{\widetilde{k}^x}  &=&  k^x \cos (\Omega s) + k^y \sin (\Omega s) \;,
\nonumber \\
{\widetilde{k}^y}  &=&  - k^x \sin (\Omega s) + k^y\cos (\Omega s) \;,
\nonumber \\
{\widetilde{k}^z}  &=&  k^z \; ,
\nonumber
\end{eqnarray}
which consists of a rotation by an angle $\Omega s$ around the $k^z$ axis.
Hence, we obtain from Eq.~(\ref{dP3a}) the  following transition rate per
momentum-space element for the emitted scalar:
\begin{eqnarray}
 \frac{
   d{\cal R}_s^{p_1 \to p_2}}{d^3{\widetilde{\bf k}}}
  &=&
\frac{
 \;\;G_{\rm eff}^{(s)2}}{(2\pi)^3{2\gamma^2\widetilde{\omega}}
     }
\int_{-\infty}^{+\infty} d\sigma\;
\exp [i(
          \Delta \mu \sigma/\gamma  \nonumber \\
          &+&\widetilde{k}^\mu X_\mu (\sigma))] \;,
\label{dG1s}
\end{eqnarray}
where
$
{\cal R}_s^{p_1 \to p_2} \equiv {d{\cal P}^{p_1 \to p_2}}/{ds} \;
$
is the transition probability per {\em laboratory} time and
\begin{equation} \label{Xmu}
X^\mu (\sigma) \equiv (\sigma\, ,\, 0\, ,\, 2R\sin (\Omega \sigma/2) \, ,\, 0) \;.
\end{equation}

In order to calculate the transition rate
\begin{equation}
{\cal R}_s^{p_1 \to p_2} \equiv
\int d^3 \widetilde{\bf k}\;
\frac{d{\cal R}_s^{p_1 \to p_2}}{d^3 \widetilde{\bf k}}\;,
\label{Gauxs}
\end{equation}
we use Eq.~(\ref{dG1s}) and obtain
\begin{equation}
{\cal R}_s^{p_1 \to p_2}
=
\frac{ G_{\rm eff}^{(s)2}}{2\;\gamma^2\;(2\pi)^3 }
\int_{- \infty}^{+\infty} d \sigma \;
e^{i\; \Delta \mu \; \sigma/\gamma}  I(\sigma)\; ,
\label{Gaux2s}
\end{equation}
where
\begin{equation}
I(\sigma)
\equiv
\int d^3 \widetilde{\bf k}
\frac{e^{i\; \widetilde{k}^\lambda X_\lambda}} {\widetilde\omega}
\label{I_a}
\end{equation}
and
$\widetilde{\omega} = \sqrt{\widetilde{\bf k}^2 + m^2}$.
In order to integrate Eq.~(\ref{I_a}), we introduce spherical coordinates
in the momenta space
$
({\widetilde{k}}\in {\rm R}^+,
 {\widetilde{\theta}}\in [0,\pi],
 {\widetilde{\phi}}\in [0,2\pi))
$,
where
$ {\widetilde{k}}^x =
{\widetilde{k}} \sin {\widetilde{\theta}} \cos {\widetilde{\phi}} $,
$ {\widetilde{k}}^y =
{\widetilde{k}} \sin {\widetilde{\theta}} \sin {\widetilde{\phi}} $,
and
$ {\widetilde{k}}^z = {\widetilde{k}} \cos {\widetilde{\theta}} $.
By doing so, we obtain
$$
I(\sigma)  = \frac{4 \pi}{|{\bf X}|}
       \int_{m}^{+\infty} d{\widetilde{\omega}}
       e^{i{\widetilde{\omega}} X^0 }
          \sin
          \left[
          \sqrt{{\widetilde{\omega}}^2 - m^2} \; | {\bf X }|
          \right]\; ,
$$
where $|{\bf X}| \equiv \sqrt{-X_i X^i\,} $. Next, by redefining the
frequency  variable as
$\widetilde{\omega} \equiv m \cosh \xi $, we obtain
$$
I(\sigma)  =
             \frac{-2\pi i m}{ |{\bf X}| } \int_{-\infty}^{+\infty} d\xi
             e^{i m (X^0 \cosh \xi + | {\bf X} | \sinh \xi) } \sinh \xi .
$$
Now, we perform the change of  variable $\xi \mapsto \eta \equiv e^{\xi}$,
leading to
\begin{eqnarray}
I(\sigma) =
\frac{ i \pi m}{| {\bf Y} |}
\int_0^{+\infty}
& &
d\eta (\eta^{-2} -1)
\exp
\left[
       \frac{i m (Y^0 + | {\bf Y} |) \eta}{2}
\right.
\nonumber \\
& &
\left.
     + \frac{i m (Y^0 - | {\bf Y} |)}{2 \eta}
     \right] \;,
\end{eqnarray}
where we have introduced a small positive regulator $\epsilon > 0$
in the integral as follows:
\begin{equation}
X^\mu \mapsto Y^\mu = (X^0 + i\epsilon, X^1, X^2, X^3)\; .
\label{ymu}
\end{equation}
(Note that $ |{\rm Re} (Y^0)| = |X^0| > |{\bf X}| = | {\bf Y} | $.)
Then, by using expressions (3.471.11) and (8.484.1) of Ref.~\cite{GR},
we obtain
\begin{equation}
I(\sigma)
 =
\frac{-2 \, \pi^2 \, i\, \, m \,{\rm sign}(\sigma)  }{\sqrt{Y_\mu Y^\mu \,}}
H_1^{(1)} \left( {\rm sign}(\sigma) m \sqrt{Y_\mu Y^\mu \,} \right) \;,
\label{I_afim}
\end{equation}
where
$H^{(1)}_1 (z)$ is the Hankel function of the first kind. As a result,
by making the variable change
$\sigma \mapsto \lambda \equiv - a \sigma/\gamma$
and by defining $Z^\mu \equiv (a/\gamma) Y^\mu $,
the transition rate~(\ref{Gaux2s}) can be cast in the form
\begin{equation}
{\cal R}_s^{p_1 \to p_2}
=
\frac{-iG_{\rm eff}^{(s)2} \widetilde{m}^2 a}{8 \pi\,\gamma }
\int_{- \infty}^{+\infty} d \lambda  \;
e^{- i\; \widetilde{\Delta \mu}\, \lambda }
\frac{ H_1^{(1)} ( z ) }{z}  \; ,
\label{Gfim}
\end{equation}
where we have defined
$\widetilde{m} \equiv m/a$,
$\widetilde{\Delta \mu} \equiv \Delta \mu/a$,
$\epsilon'\equiv a \epsilon/\gamma \ll 1$,
$z \equiv - \widetilde{m} \gamma {\rm sign}(\lambda)
\sqrt{Z_\lambda Z^\lambda \,}$,
and where
\begin{equation}
Z^\mu
       = (-\lambda + i \epsilon',
          0,
          -(2Ra/\gamma) \sin (\Omega \lambda \gamma/2a) ,
          0).
\label{zmu}
\end{equation}
Eq.~(\ref{Gfim}) is our general formula for the transition rate per
laboratory time.

In the physically interesting regime, where $\widetilde{m} \ll 1$ it can
be integrated using the following expansion for the
Hankel function~\cite{GR}:
\begin{equation}
H_1^{(1)} (z) \approx -\frac{2 i }{\pi z}
                         + {\cal O} (z\ln z)
\; \; {\rm for} \;\; |z| \ll 1 \; .
\label{approx1S}
\end{equation}
We note that for large enough
$|\lambda |$, $|z|>1$,
Eq.~(\ref{approx1S}) ceases to be a good approximation. [For instance, for
$\gamma^2 \gg 1/\widetilde{m} \gg 1$,
we have that $|z|>1$ for
$|\lambda | \geq 1/\sqrt{12 \widetilde{m} \,}$,
while for $1/\widetilde{m} \gg \gamma^2 \gg 1$,
we have that
$|z|>1$ for $|\lambda | \geq 1/ (\gamma \widetilde{m} )$.]
Notwithstanding, this is not important because the error committed
in this region is small to affect the final result provided that
$\widetilde{m} \ll 1$. Hence we write Eq.~(\ref{Gfim}) for
$\widetilde{m} \ll 1$ in the form
\begin{equation}
{\cal R}_s^{p_1 \to p_2}
\approx
\frac{-G_{\rm eff}^{(s)2}  a }{4 \pi^2 \gamma^3}
\int_{- \infty}^{+\infty} d \lambda \;
\frac{e^{- i\; \widetilde{\Delta \mu}\, \lambda }}{(Z^\lambda Z_\lambda)} \;,
\label{Gfimaprox1}
\end{equation}
where
\begin{equation}
Z_\lambda Z^\lambda  =
(\lambda-i\epsilon')^2 - (2 R a/\gamma)^2 \sin^2 (\Omega \lambda \gamma/2a)\,.
\label{W}
\end{equation}
In order to solve this integral, we expand $Z^\lambda Z_\lambda$ for
relativistic swirling particles~\cite{T}, i.e., $\gamma \gg 1$
(recall that $R= v^2 \gamma^2/ a$, $\Omega = a/(v \gamma^2)$, and
$v=\sqrt{1-\gamma^{-2}}$):
\begin{eqnarray}
Z_\lambda Z^\lambda
& \approx &
\frac{1}{12\, \gamma^{2}}
(\lambda + i \sqrt{3} A_+)(\lambda + i \sqrt{3} A_-)
(\lambda - i \sqrt{3} B_+)
\nonumber \\
& &\times
(\lambda - i \sqrt{3} B_-) \;,
\label{zexpansion}
\end{eqnarray}
where
$$
A\mp \equiv 1 \mp \sqrt{1+ 2 \widetilde{\epsilon} /\sqrt{3}}
$$
and
$$
B\mp \equiv 1 \mp \sqrt{1- 2 \widetilde{\epsilon} /\sqrt{3}}
$$
with
$\widetilde{\epsilon} \ll 1$. For $|\lambda| \gtrsim 2 v \gamma$,
where the expansion ceases to be a good approximation, the integral
contributes very little again and, thus, will not have any major
influence in the final result. Thus, the integral in Eq.~(\ref{Gfimaprox1})
can be rewritten in the complex plane:
\begin{equation}
{\cal R}_s^{p_1 \to p_2}
\approx
\frac{-G_{\rm eff}^{(s)2}  a }{4 \pi^2 \gamma^3}
\oint_{C} d \lambda \;
\frac{e^{- i\; \widetilde{\Delta \mu}\, \lambda }}{Z^\lambda Z_\lambda}\;,
\label{Gfimaprox2s}
\end{equation}
\begin{figure}
\includegraphics[width=0.3\textwidth]{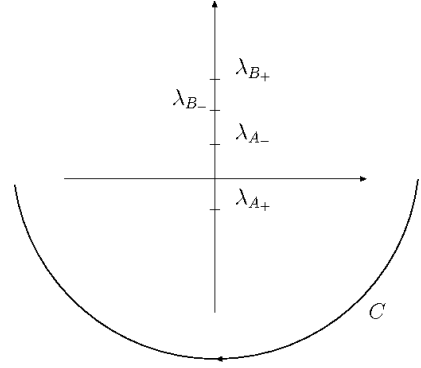}
\caption{The integration path in complex plane.
The $\lambda_{A\pm,B\pm}$ refers to the
poles for each of the terms in parentheses of the Eq.~(\ref{zexpansion}).}
\label{residuepath}
\end{figure}

\noindent where the complex integration path $C$ is clockwise oriented
($C \equiv (-L,L) \cup \{L \; e^{i \theta},
\theta \in [-\pi, 0] \}\,,L\to \infty$)
as shown in Fig.~\ref{residuepath}. Eq.~(\ref{Gfimaprox2s}) can be performed,
then, by using Cauchy's residue theorem leading to
\begin{equation}
{\cal R}_s^{p_1 \to p_2}
\approx
\frac{
      G_{\rm eff}^{(s)2}  a}
     {
      8 \sqrt{3}{\pi }\gamma
      }\,e^{{- 2\sqrt{3}\widetilde{\Delta \mu}}}\;,
\label{taxadecaimentoforteinercial}
\end{equation}
which is valid for $\widetilde{m} \ll 1$ and $\gamma \gg 1$.

\subsection{\label{fermionic case} Fermionic case}

Next, let us compute the transition rate associated with the
fermionic $f_1-\bar{f}_2$ emission of process (\ref{p1p2b}).
After performing the variable changes~(\ref{CCGs}) and~(\ref{Gamma/k1k2s}),
one obtains from Eq.~(\ref{dP3}),
\begin{eqnarray}
&& \frac{
   d{\cal R}_f^{p_1 \to p_2}}{d^3{\widetilde{\bf k}}_1
   d^3{\widetilde{\bf k}}_2
     }
 \; = \;
\frac{
 2\;G_{\rm eff}^{(f)2}}{(2\pi)^6{\widetilde{\omega}}_1{\widetilde{\omega}}_2
     }
\int_{-\infty}^{+\infty} d\sigma\;
          \exp [i(
          \Delta \mu \sigma/\gamma  \nonumber\\
         && + (\widetilde{k}_1 + \widetilde{k}_2)^\mu X_\mu (\sigma) ) ] [
\, R^2 \, \Omega^2
(
 {\widetilde{\omega}}_1 {\widetilde{\omega}}_2 -
 {\widetilde{k}}^z_1 {\widetilde{k}}^z_2
)
\cos (\, \Omega \sigma\,  )  \nonumber \\
&&  - R^2 \Omega^2
(
 {\widetilde{k}}^x_1 {\widetilde{k}}^x_2 -
 {\widetilde{k}}^y_1 {\widetilde{k}}^y_2
) + ({\widetilde{\omega}}_1 {\widetilde{\omega}}_2
  + {\widetilde{\bf k}}_1 \cdot {\widetilde{\bf k}}_2)
\nonumber \\
& &
-\,  2 \, R \, \Omega \, ({\widetilde{\omega}}_1 {\widetilde{k}}^y_2
+{\widetilde{\omega}}_2 {\widetilde{k}}^y_1)\cos (\, \Omega \sigma/2\,  )
\nonumber\\
&&
+ \, 2\,  i\,  R \, \Omega \, ({\widetilde{\bf k}_1}
\times {\widetilde{\bf k}_2})^x\sin (\, \Omega  \sigma/2 \, )
 \nonumber \\
& &
- \, i \, R^2 \, \Omega^2\,
(
{\widetilde{\omega}}_1 {\widetilde{k}}^z_2 -
{\widetilde{\omega}}_2 {\widetilde{k}}^z_1
)
\sin (\, \Omega \sigma \, )] \; ,
\label{dG1}
\end{eqnarray}
which is the laboratory transition rate per momentum-space element
associated with each emitted fermion and $X^\mu$ is given in Eq.~(\ref{Xmu}).
By integrating over all momenta, the transition rate can be rewritten
in a more convenient form as
\begin{equation}
{\cal R}^{p_1 \to p_2}_f
=
\frac{2\; G_{\rm eff}^{(f)2}}{(2\pi)^6 }
\int_{- \infty}^{+\infty} d \sigma \;
e^{i\; \Delta \mu \; \sigma/\gamma}
G_{\mu \nu} A^{\mu \nu}
\label{Gaux2}
\end{equation}
with
\begin{equation}
G_{\mu \nu}
\equiv
- \frac{\partial I_1}{\partial X^\mu}
  \frac{\partial I_2}{\partial X^\nu}
\label{G}
\end{equation}
and
\begin{equation}
I_l (\sigma)
\equiv
\int d^3 \widetilde{\bf k}_l
\frac{e^{i\; \widetilde{\bf k}_l^\lambda X_\lambda}}
{\widetilde\omega_l} \;,
\label{I_af}
\end{equation}
where the index $l=1,2$ is used to distinguish the fermion in the
final state to which we are referring and
$\widetilde{\omega}_l = \sqrt{\widetilde{\bf k}_l^2 + m_l^2}$. Also
\begin{widetext}
\begin{equation}
A_{\mu \nu} =
\left[
 \begin{array}{cccc}
 1+R^2\Omega^2\cos(\Omega\sigma) & 0 &
             -2R\Omega\cos(\Omega\sigma/2) & -iR^2\Omega^2\sin(\Omega\sigma)\\
 0 & 1-R^2\Omega^2 &
             0 & 0\\
 -2R\Omega \cos (\Omega\sigma/2) & 0 &
             1+R^2\Omega^2 & 2iR\Omega \sin(\Omega\sigma/2)\\
 iR^2\Omega^2\sin(\Omega\sigma) & 0 &
            -2iR\Omega\sin(\Omega\sigma/2) & 1-R^2\Omega^2\cos(\Omega\sigma)
 \end{array}
\right] \; .
\label{A}
\end{equation}
In order to compute Eq.~(\ref{I_af}), we introduce
spherical coordinates in the momenta space
$
({\widetilde{k}}_l\in {\rm R}^+,
 {\widetilde{\theta}}_l\in [0,\pi],
 {\widetilde{\phi}}_l\in [0,2\pi))
$,
where
$
{\widetilde{k}}_l^x
=
{\widetilde{k}}_l \sin {\widetilde{\theta}_l} \cos {\widetilde{\phi}_l}
$,
$
{\widetilde{k}}_l^y
=
{\widetilde{k}}_l \sin {\widetilde{\theta}_l} \sin {\widetilde{\phi}_l}
$,
and
$
{\widetilde{k}}_l^z
=
{\widetilde{k}}_l \cos {\widetilde{\theta}_l}
$
and perform the same steps of the previous section which led Eq.~(\ref{I_a})
into Eq.~(\ref{I_afim}). We obtain, thus,
\begin{equation}
I_l (\sigma)
 =
\frac{-2 \pi^2 i\, m_l\, {\rm sign}(\sigma)  }{\sqrt{Y_\mu Y^\mu}}
H_1^{(1)} \left( {\rm sign}(\sigma) m_l \sqrt{Y_\mu Y^\mu} \right) \;,
\label{I_fafim}
\end{equation}
where $Y_\mu$ is defined in Eq.~(\ref{ymu}).
By introducing again $Z^\mu \equiv (a/\gamma) Y^\mu $ and the variable
$\sigma \mapsto \lambda \equiv - a \sigma/\gamma$, the
transition rate~(\ref{Gaux2}) can be cast in the form
(see also expression 8.472.4 in Ref.~\cite{GR})
\begin{equation}
{\cal R}_f^{p_1 \to p_2}
=
\frac{G_{\rm eff}^{(f)2}
\widetilde{m}_1^4 \widetilde{m}_2^4 a^5\gamma^3}{8 \pi^2 }
\int_{- \infty}^{+\infty} d \lambda  \;
e^{- i\; \widetilde{\Delta \mu}\, \lambda }
Z^\mu Z^\nu A_{\mu \nu} \frac{ H_2^{(1)} ( z_1 ) }{z_1^2}\;
\frac{ H_2^{(1)} ( z_2 ) }{z_2^2}  \; ,
\label{GfimF}
\end{equation}
where
$\widetilde{m_l} \equiv m_l/a$,
$\widetilde{\Delta \mu} \equiv \Delta \mu/a$,
$\epsilon'\equiv a \epsilon/\gamma \ll 1$,
$z_l \equiv - \widetilde{m}_l \gamma {\rm sign}(\lambda)
\sqrt{Z_\lambda Z^\lambda \,}$
and $Z_\mu$ is given in Eq.~(\ref{zmu})
with
\begin{equation}
A_{\mu \nu} =
\left[
 \begin{array}{cccc}
 1+R^2\Omega^2\cos(\Omega\gamma\lambda/a) & 0 &
                              -2R\Omega\cos(\Omega\gamma\lambda/2a) &
                                     iR^2\Omega^2\sin(\Omega\gamma\lambda/a)\\
 0 & 1-R^2\Omega^2 & 0 & 0\\
 -2R\Omega \cos (\Omega \gamma \lambda/2 a) & 0 &
                              1+R^2\Omega^2 &
                                     -2iR\Omega\sin(\Omega\gamma\lambda/2a)\\
 -iR^2\Omega^2\sin(\Omega\gamma\lambda/a) & 0 &
                              2iR\Omega\sin(\Omega\gamma\lambda/2a) &
                                  1-R^2\Omega^2 \cos(\Omega \gamma \lambda/a)
 \end{array}
\right] \; .
\label{Afin}
\end{equation}
\end{widetext}
This is our general expression for the laboratory reaction rate
associated with the process~(\ref{p1p2b}).

Next,  we cast Eq.~(\ref{GfimF}) in a simpler form in the regime
where $\widetilde{m_l} \ll 1$. For this purpose we use the
expansion for the Hankel function~\cite{GR}
\begin{equation}
H_2^{(1)} (z_l) \approx -\frac{4 i }{\pi z_l^2} -\frac{i}{\pi}
                         + {\cal O} (z_l^2 \ln z_l)
\; \; {\rm for} \;\; |z_l| \ll 1
\label{approx1}
\end{equation}
and use a similar reasoning presented below Eq.~(\ref{approx1S})
[with the suitable identifications $z \to z_l$ and
$\widetilde m \to \widetilde m_l$] to obtain
\begin{eqnarray}
 {\cal R}_f^{p_1 \to p_2}
 \approx & &
\frac{-G_{\rm eff}^{(f)2}  a^5 }{8 \pi^4 \gamma}
\int_{- \infty}^{+\infty} d \lambda \;
e^{- i\; \widetilde{\Delta \mu}\, \lambda }
\frac{Z^\mu Z^\nu A_{(\mu \nu)}}{(Z^\lambda Z_\lambda)^2}
\nonumber \\
& &
\times
\left(
 \frac{16 }{\gamma^4 (Z_\lambda Z^\lambda)^2}
+\frac{4(\widetilde{m}_1^2 + \widetilde{m}_2^2)}{\gamma^2 Z_\lambda Z^\lambda}
\right)\; ,
\label{Gfimaprox1F}
\end{eqnarray}
where $Z^\lambda Z_\lambda$ is given in Eq.~(\ref{zexpansion})
for  $\gamma \gg 1$ (recall that $R= v^2 \gamma^2/ a$,
$\Omega = a/(v \gamma^2)$, and $v=\sqrt{1-\gamma^{-2}}$).
As in the scalar case, the integral above is performed in the complex
plane along the path given in Fig.~\ref{residuepath}:
\begin{eqnarray}
{\cal R}_f^{p_1 \to p_2}
\approx & &
\frac{-G_{\rm eff}^{(f)2}  a^5 }{8 \pi^4 \gamma}
\oint_{C} d \lambda \;
e^{- i\; \widetilde{\Delta \mu}\, \lambda  }
\frac{Z^\mu Z^\nu A_{(\mu \nu)}}{(Z^\lambda Z_\lambda)^2}
\nonumber \\
& &
\times
\left(
 \frac{16}{\gamma^4 (Z_\lambda Z^\lambda)^2}
+\frac{4(\widetilde{m}_1^2 + \widetilde{m}_2^2)}{\gamma^2 Z_\lambda Z^\lambda}
\right)\; .
\label{Gfimaprox2}
\end{eqnarray}
Then, by using the Cauchy's residue theorem, we obtain for
$\widetilde{m}_1,\widetilde{m}_2 \ll 1$ and $\gamma \gg 1$
\begin{eqnarray}
&&{\cal R}_f^{p_1 \to p_2}
\approx
\frac{G_{\rm eff}^{(f)2}  a^{5}
\exp({- 2\sqrt{3}\widetilde{\Delta \mu}})}{1728 {\pi }^3
{\mathbf\gamma}}[49 \sqrt{3} + 102\widetilde{\Delta \mu} \nonumber\\
 && + 30\sqrt{3}\widetilde{\Delta \mu}^2
 + 12\,\widetilde{\Delta \mu}^3 -39\,\sqrt{3}\,( {\widetilde{m}_1}^2
+ {\widetilde{m}_2}^2 )\nonumber \\
&& - 90\,\widetilde{\Delta \mu}\,({\widetilde{m}_1}^2 + {\widetilde{m}_2}^2)
  - 36\,\sqrt{3}\,\widetilde{\Delta \mu}^2\,({\widetilde{m}_1}^2
+ {\widetilde{m}_2}^2) ].
\label{taxadecaimentofracoinercial}
\end{eqnarray}
This is easy to note that Eq.~(\ref{taxadecaimentofracoinercial})
is positive definite and decreases as $\widetilde{m}_1$ and
$\widetilde{m}_2$ increase. It is not difficult to show, as well,
that it also decreases as  $\widetilde{\Delta \mu}$ increases, as expected.

As a check of our approach, let us use our formulas to analyze the usual
$\beta$-decay:
$n^0 \to p^+ \, e^-\, \bar \nu$. The mean proper lifetime of inertial
{\em neutrons} is $887$ s~\cite{PDG}. Thus,
$
{\cal R}^{n\to p}_{in} \equiv
{\cal R}^{n\to p}_f(\Omega\to 0)
= \hbar/887$~s
leads to
\begin{equation}
{\cal R}^{n\to p}_{in}
= 5.46 \times 10^{-3} \; G_F^2 \; {\rm MeV}^5\;,\label{GEFF}
\end{equation}
where $G_F \equiv 1.166\times 10^{-5}\;{\rm GeV}^{-2}$ is the Fermi
coupling constant~\cite{PDG}. Clearly, we cannot use our
expression~(\ref{Gfimaprox1F}) to calculate the reaction rate of
the $\beta$-decay, since it is not valid for inertial neutrons.
However, ${\cal R}_{in}^{n \to p}$  can be derived in this case
directly from Eq.~(\ref{Gaux2}) by making $\Omega = 0$ in Eq.~(\ref{dG1}).
This is achieved by a change of the momentum variables as shown in
Eq.~(\ref{Gamma/k1k2s}). After performing the corresponding integrations
in the angular coordinates and in $\widetilde{\omega}_e$, we obtain
\begin{eqnarray}
{\cal R}^{n\to p}_{in} & = &
\frac{G_{pn}^2 }{\pi^3}
\int_0^{\Delta \mu -m_e}
d{\widetilde{\omega}}_\nu\;{\widetilde{\omega}}_\nu^2
\left(
\Delta \mu  - \widetilde{\omega}_\nu
\right)
\nonumber \\
& \times &
\sqrt{
\left(
\Delta \mu - \widetilde{\omega}_\nu
\right)^2- m_e^2}\;,
\label{dGIN2}
\end{eqnarray}
where we have assumed massless neutrinos,
$m_\nu = 0$, and $G_{pn} \equiv G^{(f)}_{\rm eff}$.
By evaluating numerically Eq.~(\ref{dGIN2})  with
$m_e= 0.511 \; {\rm MeV}$
and
$\Delta \mu = (m_n-m_p)= 1.29 \;{\rm MeV}$, we obtain
$$
{\cal R}^{n\to p}_{in}\;=\;
1.81 \times 10^{-3} \;G_{pn}^2 \; {\rm MeV}^5 \;.
$$
This is to be compared with Eq.~(\ref{GEFF}),
where $G_{pn}$ is to be identified with $G_F$.
The reason why both results are not identical can be traced back to
the fact that the nucleons are treated here semiclassically and have
only approximately the same kinetic energy content:
the no-recoil condition only models approximately the real physical
situation. Notwithstanding, this suffices for our present purposes.

\setcounter{equation}{0}

\section{\label{power} Emitted power}

\subsection{\label{scalar case2} Scalar case}

Next, we calculate the radiated power
\begin{equation}
W^{p_1 \to p_2}_s
\equiv
\int d^3 \widetilde{\bf k}
\; \widetilde{\omega}
\frac{
      d {\cal R}_s^{p_1 \to p_2}
     }{
     d^3 \widetilde{\bf k}
     }
\label{WauxS}
\end{equation}
associated with the emitted scalars as measured in the
laboratory frame. Eq.~(\ref{WauxS}) can be rewritten as
\begin{equation}
W^{p_1 \to p_2}_s
=
\frac{ G_{\rm eff}^{(s)2}}{(2\pi)^3 \,\gamma}
\int_{- \infty}^{+\infty} d \sigma \;
e^{i\; \Delta \mu \; \sigma/\gamma}
J(\sigma) \; ,
\label{WauxS2}
\end{equation}
where
\begin{equation}
J(\sigma)
\equiv
\int d^3 \widetilde{\bf k}
e^{i\; \widetilde{\bf k}^\lambda X_\lambda}
\label{J_1}
\end{equation}
and $X^\mu$ is given in Eq.~(\ref{Xmu}). In order to integrate $J(\sigma)$,
we follow closely the approach, which drove Eq.~(\ref{I_a}) into
Eq.~(\ref{I_afim}):
\begin{equation}
J(\sigma) =
\frac{2 \, \pi^2  \, m^2\, Y_0  }{Y_\mu Y^\mu }
H_2^{(1)} \left( {\rm sign}(\sigma) m \sqrt{Y_\mu Y^\mu \,} \right) \;,
\label{J_1fim}
\end{equation}
where $Y^\mu$ is given in Eq~(\ref{ymu}).
Now, by introducing again $\sigma \mapsto \lambda \equiv - a \sigma/\gamma$
and $Z^\mu \equiv (a/\gamma) Y^\mu $,
Eq.~(\ref{WauxS2}) can be cast in the form
\begin{equation}
W^{p_1 \to p_2}_{s}
=
\frac{
      G_{\rm eff}^{(s)2} \widetilde{m}^4 \gamma a^4
      }{
      8 \pi^2
      }
\int_{- \infty}^{+\infty} d \lambda \;\lambda\;
e^{- i\; \widetilde{\Delta \mu}\, \lambda  }
\frac{ H_2^{(1)} ( z ) }{z^2} \;,
\label{WfimS}
\end{equation}
where $z$  can be found below Eq.~(\ref{Gfim}) and $Z^\mu$ is given
in Eq.~(\ref{zmu}). Eq.~(\ref{WfimS}) is the general expression for
the radiated power associated with the emitted scalars.

The expression above can be simplified in the limit
$\widetilde{m} \ll 1$. For this purpose we use the expansion
(see Ref.~\cite{GR})
\begin{equation}
H_2^{(1)}(z)\approx -\frac{4 i }{\pi z^2} + {\cal O} (z^0)
\label{approx2S}
\end{equation}
for $ |z| \ll 1 $. Then, by letting Eq.~(\ref{approx2S}) in
Eq.~(\ref{WfimS}), we can perform the remaining integral in the
complex plane along the path of Fig.~\ref{residuepath} to obtain
the emitted power in the regime
$\widetilde{m}\ll 1$ and $\gamma \gg 1$
\begin{equation}
W^{p_1 \to p_2}_s
\approx
\frac{G_{\rm eff}^{(s)2} a^2 e^{-2\sqrt{3} \widetilde{\Delta \mu}}}{12 \pi }
\left(
      1+ \frac{\sqrt{3}}{2} \widetilde{\Delta \mu}
\right) \;.
\label{w1s}
\end{equation}
This is in agreement with the expression obtained by Ginzburg and
Zharkov~\cite{GZ} (see also Ref.~\cite{CHM}) in the due limit, i.e.,
$\widetilde{\Delta \mu} \to 0$.

\subsection{\label{fermionic case2} Fermionic case}

Further, we calculate the radiated power as measured by observers at rest
in the laboratory frame associated with each fermion $l=1, 2$:
\begin{equation}
W^{p_1 \to p_2}_{f(l)}
\equiv
\int d^3 \widetilde{\bf k}_1
\int d^3 \widetilde{\bf k}_2
\; \widetilde{\omega}_l
\frac{
      d {\cal R}_f^{p_1 \to p_2}
     }{
     d^3 \widetilde{\bf k}_1 d^3 \widetilde{\bf k}_2
     }\;.
\label{Waux}
\end{equation}
Eq.~(\ref{Waux}) can be rewritten as
\begin{equation}
W^{p_1 \to p_2}_{f(1)}
=
\frac{2\; G_{\rm eff}^{(f)2}}{(2\pi)^6 }
\int_{- \infty}^{+\infty} d \sigma \;
e^{i\; \Delta \mu\; \sigma/\gamma}
H_{\mu \nu} A^{\mu \nu} \; ,
\label{Waux2}
\end{equation}
where we have chosen (with no loss of generality) $l=1$,
i.e., we are computing the radiated power associated with the
fermion with mass $m_1$. Here
\begin{equation}
H_{\mu \nu}
\equiv
- \frac{\partial J_1}{\partial X^\mu}
  \frac{\partial I_2}{\partial X^\nu} \;,
\label{H}
\end{equation}
where
\begin{equation}
J_1 (\sigma) \equiv
\int d^3 \widetilde{\bf k}_1
e^{i\; \widetilde{\bf k_1}^\lambda X_\lambda} \;,
\label{J_1f}
\end{equation}
and $I_2$ is given in Eq.~(\ref{I_fafim}) with $l=2$. The result of
Eq.~(\ref{J_1f}):
\begin{equation}
J_1 (\sigma)=
\frac{2 \, \pi^2  \, m_1^2\, Y_0  }{Y_\mu Y^\mu }
H_2^{(1)} \left( {\rm sign}(\sigma) m_1 \sqrt{Y_\mu Y^\mu \,} \right)
\label{J_1fimF}
\end{equation}
is obtained by inspection after comparing Eq.~(\ref{J_1}) with
Eq.~(\ref{J_1f}) and Eq.~(\ref{J_1fim}) with Eq.~(\ref{J_1fimF}),
respectively, where $z_l$ is defined below Eq.~(\ref{GfimF}),
and $Z^\mu$ and $A_{\mu \nu}$ are given by Eqs.~(\ref{zmu})
and~(\ref{Afin}), respectively. By letting Eqs.~(\ref{I_fafim})
(with $l=2$) and~(\ref{J_1fimF}) in Eq.~(\ref{H}), we rewrite the
emitted power~(\ref{Waux2}) in the form
\begin{eqnarray}
W^{p_1 \to p_2}_{f(1)}
& = &
\frac{
      G_{\rm eff}^{(f)2} \widetilde{m}_1^4 \widetilde{m}_2^4  a^6 \,i
      }{
      8 \pi^2 \gamma^{-2}
      }
\int_{- \infty}^{+\infty}\!\!\!\!\! d \lambda \;
e^{- i\; \widetilde{\Delta \mu}\, \lambda  }
\frac{ H_2^{(1)} ( z_2 ) }{z_2^2}
\nonumber \\
&& \!\!\!\!\!\!\!\!
\times
\left[ \frac{ H_3^{(1)} ( z_1 ) }{z_1^3}
      \widetilde{m}_1^2 \gamma^2 Z^0Z^\mu Z^\nu A_{(\mu \nu)}
\right.
\nonumber \\
&&
\left.
      - \frac{ H_2^{(1)} ( z_1 ) }{z_1^2}\; \eta^{0 \mu} Z^\nu A_{\mu \nu}
\right] \! .
\label{WfimF}
\end{eqnarray}
This is our general formula for the total emitted power associated
with the fermion $l=1$.

In the limit $\widetilde{m}_l \ll 1$, we can rewrite
$W^{p_1 \to p_2}_{f(1)}$ by using the
expansions~(\ref{approx1}) and (see Ref.~\cite{GR})
\begin{equation}
H_3^{(1)}(z_l)\approx -\frac{16 i }{\pi z_l^3} -\frac{2i}{\pi z_l}
                      -\frac{z_l i}{4 \pi}
                      + {\cal O} (z_l^3 \ln z_l) \;,
\label{approx2}
\end{equation}
for $ |z_l| \ll 1 $. Thus, by letting Eqs.~(\ref{approx1}) and
(\ref{approx2}) in Eq.~(\ref{WfimF}), we can perform the remaining
integral in the complex plane along the same path shown in
Fig.~\ref{residuepath} and obtain the emitted power for
$\widetilde{m}_1,\widetilde{m}_2 \ll 1$ and $\gamma \gg 1$:
\begin{eqnarray}
&&
\!\!\!\!
W^{p_1 \to p_2}_{f(1)}
\!\approx\!
\frac{G_{\rm eff}^{(f)2} a^6 e^{-2\sqrt{3}\;
\widetilde{\Delta \mu}}}{3456 \pi^3 }
\left[
      320 + 241\sqrt{3} \; \widetilde{\Delta \mu} \right. \nonumber \\
      &&\!\!\!\left. + 246 \widetilde{\Delta \mu}^2
      + 46\sqrt{3}\;\widetilde{\Delta \mu}^3
      + 12 \widetilde{\Delta \mu}^4
      - 48 ( \widetilde{m}_1^2  + 5\widetilde{m}_2^2 ) \right.\nonumber \\
&&\!\!\!\left. - 3 \sqrt{3}
\widetilde{\Delta \mu} (17 \widetilde{m}_1^2 + 65 \widetilde{m}_2^2)
  - 18 \widetilde{\Delta \mu}^2
( 5 \widetilde{m}_1^2 + 13 \widetilde{m}_2^2 )\right.\nonumber\\
&&\!\!\!\left.  - 24 \sqrt{3} \widetilde{\Delta \mu}^3
(\widetilde{m}_1^2+ 2 \widetilde{m}_2^2)
\right]\; .
\label{w1}
\end{eqnarray}
Clearly, $W^{p_1 \to p_2}_{f (2)}$ is obtained by exchanging
$m_1\longleftrightarrow m_2$ in Eq.~(\ref{w1}).
This is important to note that Eq.~(\ref{w1}) is positive definite and
decreases as $\widetilde{m}_1$, $\widetilde{m}_2$ and
$\widetilde{\Delta \mu}$ increase, as expected.


As a consistency check of our Eq.~(\ref{w1}), let us apply it to
analyze the emission of neutrino- antineutrino pairs from
accelerated electrons: $ e^- \to e^- \; \nu_e \; \bar \nu_e $ and
compare the results in the proper limit with the ones in the
literature  obtained  when the electrons are quantized in a
background magnetic field (see, e.g., \cite{BK}-\cite{LP} and
references therein). (Comprehensive accounts on $\gamma$ and
$\nu-\bar\nu$ synchrotron radiation emitted from electrons in
magnetic fields can be found, e.g., in Ref.~\cite{BLP} and Sec.~6.1
of Ref.~\cite{BKS}, respectively, and in Ref.~\cite{nikishov85}.) The fact that we
are assuming that our source is under the influence of a
gravitational force rather than being immersed in an electromagnetic
field is not relevant in this particular case, since the neutrinos
are chargeless (An account on the degradation of the neutrinos' energy in strong
magnetic fields can be found in Ref.~\cite{GKMV}.). As a consequence, our results and the aforementioned
ones in the literature are expected to be in good agreement in the
no-recoil regime $\chi \equiv a/m_e \ll 1$. The {\it total} radiated
power of neutrino-antineutrino pairs from circularly moving
electrons in a constant magnetic field $B$ with proper acceleration
$a=\gamma e B/m_e \ll m_e$ (no-recoil condition) can be easily
calculated $ {W}^{LP}_{\nu {\bar \nu}} =
{5\;(2\;C_V^2+23\;C_A^2)G_F^2\;m_e^6\chi^6}/(108\pi^3) $ from the
differential emission rate given, e.g., in Ref.~\cite{LP} or
Ref.~\cite{BKS}. Then (see Eq.~(6.6) in Ref.~\cite{VM3}),
$$
{W}^{LP}_{\nu {\bar \nu}}
=
1.1\times 10^{-2}\;G_F^2\;a^6 \;,
$$
where we have used that the vector and axial contributions to the
electric current are $C_V^2=0.93$ and $C_A^2=0.25$~\cite{KLY},
respectively, and $\chi \equiv a/m_e \ll 1$. This is to be compared
with the result obtained from Eq.~(\ref{w1}) by defining
$G_{e\nu} \equiv G_{\rm eff}^{(f)}$ and assuming $ \Delta \mu =m_\nu =0$:
$$
{W}_{\nu {\bar \nu}} \approx
1 \times 10^{-2} G_{e\nu}^2 \;a^6 \;,
$$
where $G_{e\nu}$ is the corresponding effective coupling constant,
which is to be associated with the Fermi constant.

\setcounter{equation}{0}

\section{\label{proton} Proton decay}

Now, let us use our results to analyze the weak and strong proton
decay processes~(\ref{pne+nu2}) and~(\ref{pnpi+}), respectively.
Our formulas~(\ref{GfimF}) and~(\ref{WfimF}), and (\ref{Gfim})
and~(\ref{WfimS}) associated with the weak and strong reactions,
respectively, are quite general although cumbersome to compute.
Happily, we can use the much more friendly
ones:~(\ref{taxadecaimentofracoinercial}) and~(\ref{w1}),
and~(\ref{taxadecaimentoforteinercial}) and~(\ref{w1s}),
which  are valid in the physical regime where processes~(\ref{pne+nu2})
and~(\ref{pnpi+}) are more important. In the region where
\begin{equation}
m_{e} \ll a \ll m_{\pi}\
\label{afraco}
\end{equation}
with $m_\pi$ being the $\pi^+$ mass, the reaction~(\ref{pne+nu2})
has a non-negligible rate and dominates over the reaction~(\ref{pnpi+}).
In this case, Eqs.~(\ref{taxadecaimentofracoinercial}) and~(\ref{w1})
can be used provided that $\gamma \gg 1$. Now, in the region where
\begin{equation}
m_{\pi} \ll a \ll m_{p} \;,
\label{aforte}
\end{equation}
the reaction~(\ref{pne+nu2}) is overcome by the strong one~(\ref{pnpi+}),
in which case Eqs.~(\ref{taxadecaimentoforteinercial}) and~(\ref{w1s})
should be used. Next, we look for orbits around compact object, where
conditions~(\ref{afraco}) and~(\ref{aforte}) are verified.
\begin{figure}
\epsfig{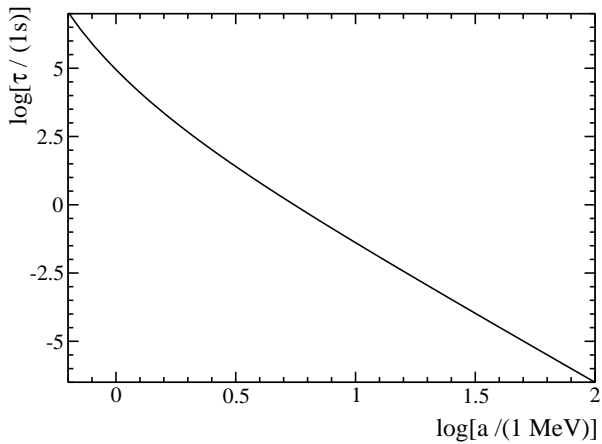}
\caption{\label{tempofracog100}
The proton mean {\em proper} lifetime $\tau$ associated with
process~(\ref{pne+nu2}) is plotted as a function of its proper
acceleration $a$, where $m_e \lesssim a \lesssim m_\pi$. }
\end{figure}

Let us begin by rewriting the proper acceleration of the proton in
Minkowski space,
$a = R \Omega^2 \gamma^2$ [see below Eq.~(\ref{UACC})], in the form
\begin{equation}
a = \Omega \gamma \sqrt{\gamma^2 -1} \; .
\label{accelerationimproved}
\end{equation}
Now, we use General Relativity to obtain the proton's energy per
mass $E/m_p$ and angular velocity $d \phi/d\tau_s$ as calculated by
a {\em static} observer lying at rest at the same radius of the
particle orbit around a compact object with mass $M$. $E/m_p$ and $d
\phi/d \tau_s$ are to be identified with $\gamma$ and $\Omega$ in
Eq.~(\ref{accelerationimproved}), respectively, to obtain the proper
acceleration $a$. Once we have $a$ and $\gamma$, we use
Eqs.~(\ref{taxadecaimentofracoinercial}), (\ref{w1}),
(\ref{taxadecaimentoforteinercial}) and~(\ref{w1s}) to calculate the
relevant decay rates and emitted powers. The results obtained in
this way should be associated with the values defined by the static
observers at the radius of the particle orbit. These ones differ
from the reaction rates and emitted powers as measured at infinity
by red-shift factors. In order to obtain (i)~the reaction rates and
(ii)~the emitted powers at infinity from the ones measured by the
static observers at the radius of the particle orbit, one should
multiply the latter ones by (i)~$\sqrt{1-2GM/r_s}$ and
(ii)~$1-2GM/r_s$, respectively. Although we can only capture with
this procedure part of the influence of the spacetime curvature, its
suitability as an approximate approach is justified by comparing the
results which it provides with the ones obtained with full curved
spacetime calculations, wherever the latter ones are available, as,
e.g., in Ref.~\cite{CHM}.

The line element external to a spherically symmetric static object with
mass $M$, which includes Schwarzschild black holes, can be written as
$$
dS^2 = \left( 1-\frac{2GM}{r} \right) dt^2 -
\left(1-\frac{2GM}{r}\right)^{-1} dr^2
                            -  d\Sigma_\Omega^2 \; ,
$$
where $d\Sigma_\Omega^2 \equiv r^2 (d\theta^2 + (\sin \theta)^2 d\phi^2)$.
According to General Relativity~\cite{W}, asymptotic observers associate an
angular velocity
$d\phi/dt_{a.o.} = \sqrt{GM/ r_s^3}$  and an energy per mass ratio
$E_{a.o.}/m = (1 - 2GM/r_s)/\sqrt{1 - 3GM/r_s}$ for particles in circular
geodesics at $r=r_s$. Thus, static observers at $r=r_s$
($\theta, \phi = {\rm const}$) associate the following corresponding values:
$$
d\phi/d\tau_s = {\sqrt{GM/ r_s^3}}/\sqrt{1-2GM/r_s}
$$
and
$$
E/m_p = \sqrt{1-2GM/r_s}/\sqrt{1 - 3GM/r_s}\; .
$$
By letting $d\phi/d\tau_s \to \Omega$ and
$E/m_p \to \gamma$, we obtain
\begin{equation}
\gamma = \sqrt{\frac{1-2GM/r_s}{1-3GM/r_s}}
\label{gammaresult}
\end{equation}
and~[see Eq.~(\ref{accelerationimproved})]
\begin{equation}
a = \frac{GM}{r_s^2 (1-3GM/r_s)} \;,
\label{accelerationresult}
\end{equation}
which will be used to evaluate Eqs.~(\ref{taxadecaimentofracoinercial}),
(\ref{w1}), (\ref{taxadecaimentoforteinercial}) and~(\ref{w1s}),
whenever $\gamma \gg 1$. We note that Eqs.~(\ref{gammaresult})
and~(\ref{accelerationresult}) are monotonic functions, which
approximate the correct values asymptotically and diverge at $r_s= 3GM$.
This is so because according to General Relativity, circular geodesic
orbits at $r_s \approx 3GM$ approximate lightlike worldlines.

At the last stable circular orbit, $r_s= 6GM$, we obtain from
Eq.~(\ref{accelerationresult}) that
$$
a/m_{e} = 3\times 10^{-16} (M_\odot/M)\;.
$$
Thus, protons around black holes in stable circular orbits $6GM <
r_s < \infty$ are not likely to decay unless the compact object is a
mini black hole with the mass of a mountain: $M \ll 10^{17}$~g [see
Eq.~(\ref{afraco})]. The fact that the smaller the black hole the
more likely that protons decay at a fixed $r_s/(GM)$ is related with
the fact that the smaller the black hole the larger the spacetime
curvature, i.e. ``gravitational field", at the same $r_s/(GM)$.
\begin{figure}
\epsfig{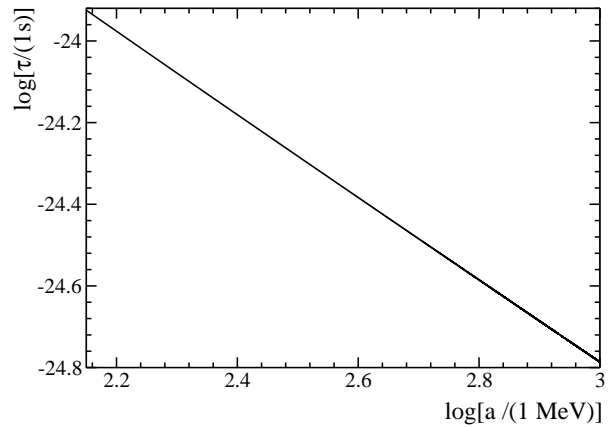}
\caption{\label{tempoforteg100}
The proton mean proper lifetime $\tau$ associated with
process~(\ref{pnpi+}) is plotted as a function of its
{\em proper} acceleration $a$, where $m_\pi \lesssim a \lesssim m_p$.
}
\end{figure}

In order to explore more realistic cases, where black holes have
some solar masses, we have to consider protons at inner circular
orbits, $3GM < r_s <6GM$, which are unstable. By defining
$r_s \equiv 3GM (1 + \delta)$ with $\delta \ll 1$ to monitor
how far from the most internal circular orbit (at $r=3GM$) the
proton is, we rewrite Eqs.~(\ref{gammaresult}) and~(\ref{accelerationresult})
as
\begin{equation}
\gamma \approx 1/\sqrt{3 \delta }
\label{accelerationresultapprox}
\end{equation}
and
\begin{equation}
a \approx 1/( 9 GM \delta ) \;.
\label{gammaresultapprox}
\end{equation}
By using Eqs.~(\ref{accelerationresultapprox}) and~(\ref{gammaresultapprox})
in Eqs.~(\ref{afraco}) and~(\ref{aforte}) we obtain
\begin{equation}
3 \times 10^{-19} (M_\odot/M) < \delta < 9 \times 10^{-17} (M_\odot/M)
\label{deltafraco}
\end{equation}
and
\begin{equation}
5 \times 10^{-20} (M_\odot/M) < \delta < 3 \times 10^{-19} (M_\odot/M)\; ,
\label{deltaforte}
\end{equation}
which are the intervals where the weak and strong processes would be
favored, respectively. Thus, free protons in circular orbits around
stellar mass black holes are likely to decay only if they are extremely
close to the most internal circular geodesic and stay there for long
enough to decay.

\begin{figure}
\epsfig{file=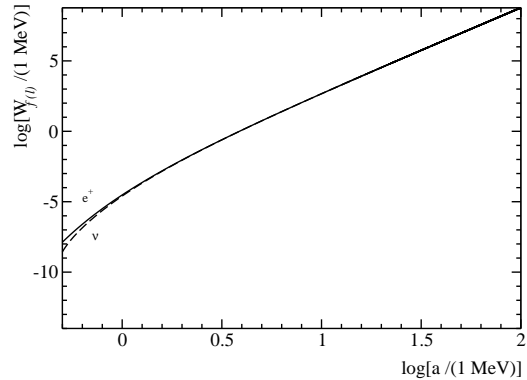,angle=270,width=0.9123\linewidth,clip=}
\caption{\label{potenciafracaenug100}
$W_{e^+}$ and $W_{\nu}$ associated with process~(\ref{pne+nu2})
are plotted as functions of the proton proper acceleration $a$
with solid and dashed lines, respectively.
}
\end{figure}

\begin{figure}[t]
\epsfig{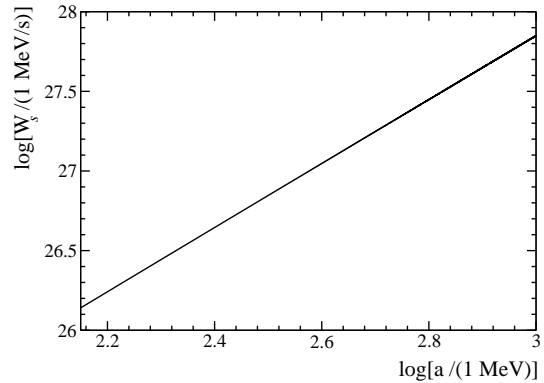}
\caption{\label{potenciafortepig100}
$W_\pi$ associated with process~(\ref{pnpi+}) is plotted as a
function of the proton proper acceleration $a$.}
\end{figure}

In Fig.~\ref{tempofracog100}, we plot
from Eq.~(\ref{taxadecaimentofracoinercial})
the proton mean {\em proper} lifetime
$\tau (a)= 1/\Gamma_f^{p\to n} $ associated with the
process~(\ref{pne+nu2}), where
$\Gamma_f^{p \to n} \equiv \gamma {\cal R}_f^{p \to n}$
is the weak transition probability per {\emph{proper}}
time and we have identified $G_{\rm eff}^{(f)} =G_{pn}$
with the Fermi coupling constant
$G_F \equiv 1.166\times 10^{-5}\;{\rm GeV}^{-2}$.
We have plotted the {\em proper} lifetime $ \tau (a)$
rather than the {\em laboratory} lifetime $t(a)$ in order
to make it easier the comparison  of this figure with
Fig.~1 in Ref.~\cite{VM3}.
In Fig.~\ref{tempoforteg100}, we plot from
Eq.~(\ref{taxadecaimentoforteinercial}) the proton mean
{\em proper} lifetime $\tau (a)= 1/\Gamma_s^{p\to n}$
associated with process~(\ref{pnpi+}), where
$\Gamma_s^{p \to n} = \gamma {\cal R}_s^{p \to n}$
is the strong transition probability per {\emph{proper}} time. Here
$G_{\rm eff}^{(s)}$ is identified with the pion-nucleon-nucleon
strong coupling constant $g_{\pi NN}$, which is written in the
Heaviside-Lorentz system as
$\sqrt{ g_{\pi NN}^2/(4\pi ) } \approx \sqrt{14}$~(see,
e.g., \cite{GZ} and~\cite{MS}).  Finally in
Figs.~\ref{potenciafracaenug100} and~\ref{potenciafortepig100},
we plot the emitted power in the form of electrons and neutrinos
as calculated from Eq.~(\ref{w1}) and in the form of pions as
calculated from~(\ref{w1s}), respectively.


\section{\label{discussion} Discussion}

The decay of accelerated protons has attracted interest for long time.
Astrophysics seems to provide suitable conditions for the observation
of the decay of accelerated protons. Cosmic ray protons with energy
$E = \gamma m_p \approx 3 \times 10^{14}$ eV under the influence
of a magnetic field $H \approx 10^{14}$ Gauss of a pulsar have proper
accelerations of $a_H = \gamma e H/m_p \approx 200$~MeV $ > m_\pi$.
For these values of $E$ and $H$, the proton are confined in a cylinder
with typical radius
$R \approx \gamma^2/a_H \approx 2\,\times \, 10^{-2} \; {\rm cm} \ll l_H $,
where $l_H$ is the typical size of the magnetic field region.
Under such conditions, protons could rapidly decay through strong
interaction before they lose most of their energy via electromagnetic
synchrotron radiation~\cite{TK}.

Here we have considered the possible weak and strong proton decays
under the influence of background gravitational fields. Reaction rates
and emitted powers were calculated. We have concluded that they are
unlikely to decay unless they orbit mini-black holes or they are pushed
to highly relativistic geodesic circular orbits (and stay there for
long enough to decay). This raises the question whether there would
exist other astrophysical sites, where the decay rate could be larger.
Perhaps the consideration of protons grazing the event horizon of black
holes or entering properly the ergosphere of Kerr black holes extracting
rotational energy from it would be worthwhile to be investigated.
Notwithstanding because these cases would involve more complicated
``trajectories'' in a genuine general relativistic context, full
quantum field theory in curved spacetime computations, rather than
our semiclassical ones, would be desirable to provide more comprehensive
results.

\begin{flushleft}
{\bf{\large Acknowledgments}}
\end{flushleft}

We are grateful to Gast\~ao Krein for providing us with reference~\cite{MS}.
D.F. and G.M. are thankful to Conselho Nacional de Desenvolvimento
Cient\'\i fico e Tecnol\'ogico for full and partial supports, respectively.
G.M. and D.V. acknowledge partial support from Funda\c c\~ao de
Amparo \`a Pesquisa do Estado de S\~ao Paulo, while D.V. is also
grateful to the US National Science Foundation for support
under the Grant No PHY-0071044 in early stages of this project.


\end{document}